\documentclass[sigconf]{acmart}

\AtBeginDocument{%
  \providecommand\BibTeX{{%
    \normalfont B\kern-0.5em{\scshape i\kern-0.25em b}\kern-0.8em\TeX}}}

\setcopyright{acmcopyright}
\copyrightyear{2019}
\acmYear{2019}
\acmDOI{10.1145/1122445.1122456}

\acmConference[CCS '19]{CCS '19: ACM Conference on Computer and Communications Security}{November 11--15, 2019}{London, UK}
\acmBooktitle{CCS '19: ACM Conference on Computer and Communications Security,
  November 11--15, 2019, London, UK}
\acmISBN{978-1-4503-5693-0/19/11}
\usepackage{algorithm}
\usepackage{algorithmic}
\usepackage{graphicx}

\begin{document}

\title{SEdroid: A Robust Android Malware Detector using Selective Ensemble Learning}

\author{Ji Wang}

\email{andrewwang@pku.edu.cn}
\affiliation{%
  \institution{Peking University}
  \city{Beijing}
  \state{China}
}

\author{Qi Jing}
\authornote{Corresponding author.}
\email{jingqi@pku.edu.cn}
\affiliation{%
  \institution{Peking University}
  \city{Beijing}
  \state{China}
}

\author{Jianbo Gao}

\email{gaojianbo@pku.edu.cn}
\affiliation{%
  \institution{Peking University}
  \city{Beijing}
  \state{China}
}

\renewcommand{\shortauthors}{Wang, et al.}

\begin{abstract}
 For the dramatic increase of Android malware and low efficiency of manual check process, deep learning methods started to be an auxiliary means for Android malware detection these years. However, these models are highly dependent on the quality of datasets, and perform unsatisfactory results when the quality of training data is not good enough. In the real world, the quality of datasets without manually check cannot be guaranteed, even Google Play may contain malicious applications, which will cause the trained model failure. To address the challenge, we propose a \emph{robust} Android malware detection approach based on selective ensemble learning, trying to provide an effective solution not that limited to the quality of datasets. The proposed model utilizes genetic algorithm to help find the best combination of the component learners and improve robustness of the model. Our results show that the proposed approach achieves a more robust performance than other approaches in the same area.
\end{abstract}



\maketitle

\section{Introduction}
 In order to address the challenges of the rapid increase of Android malwares, many deep learning approaches have been presented in recent years. Some of them are proved to be efficient and of high accuracy. Sahs et al.\cite{Sahs2012A} construct a SVM model using features consisting of Android permission and control flow graphs, which achieves a 85\% precision rate and a 95\% recall rate. Zhu et al. \cite{Zhu2017DeepFlow} introduce DBN (Deep Belief network) to this problem and achieve great results with a 95.05\% F1-score.

However, there still occurred some problems. On one hand, some researchers may find it difficult to reproduce a proven effective model in practice. On the other hand, the lack of dataset and low-quality training data may cause a model's failure. This is because of the high dependency of high-quality dataset for deep learning model. If the dataset is not good enough, the trained model may be not that good in training process and become hard to implement in application. This problem may become worse for large and high-quality dataset are commonly inaccessible for the reluctancy of holders to share the datasets for security or benefit reasons. So the robustness of the model becomes fairly important to this challenge.

\begin{figure*}[htbp]
\centering
  \includegraphics[width=0.9\linewidth]{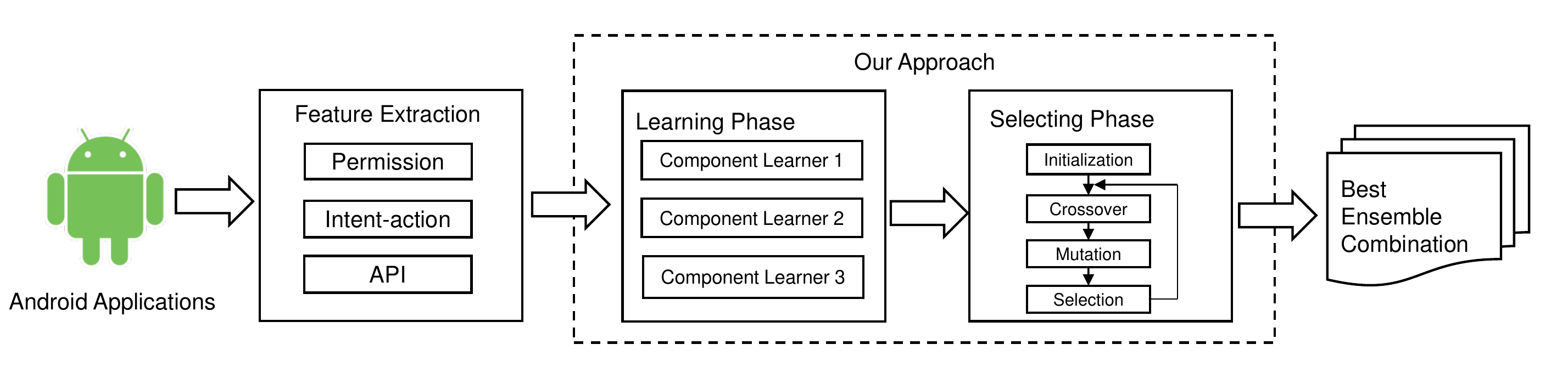}
  \caption{Overview of SEdroid\label{fig1}}
  \label{fig:overview}
\end{figure*}

\newcommand{\eg}{{\emph{e.g.}}\xspace}

In this paper, we try to exploit the architecture of selective ensemble to offer a solution with stronger generalization ability and robustness to this problem. For ensemble learning can increase memory use and computational cost, we selectively pick some of "good" ones and discard the "bad" ones. We consider DBN as component learners in our approach but we also compare it with SVM in our experiments. Since single deep neural network is easy to be over-fitting, we try to construct multiple neural networks and combine their predictions together to make a precise prediction. The main contribution can be summarized as follows:
\begin{itemize}
\item We propose a more robust Android malware detection approach based on selective ensemble learning and genetic algorithm, which is called SEdroid. Through comparative experiment, SEdroid showed a more robust and preeminent capability in Android malware detection with a 98.3\% precision rate and a 98.1\% malware recall rate.
\item The first time that selective ensemble learning is applied in Android malware detection. The algorithm we designed based on genetic algorithm considers both accuracy and diversity of the ensemble, which makes it easier to find the optimal ensemble combination and gives the model stronger generalization ability and robustness. 
\end{itemize}

\section{Approach \label{section3}}

The general architecture is shown in Figure~\ref{fig:overview}. First, we extract 3 types of features from Android application packages (.apk file) and vectorize them, which are android-permission, intent-action and API calls. Second, vectorized features are sampled to construct component neural networks using bootstrap sampling, which means that we randomly sample the data with replacement. Third, we designed a kind of genetic algorithm that consider both accuracy and diversity of the ensemble to find the best ensemble combination. Finally, the selected component neural networks are combined to make up an ensemble via majority voting, which is used to detect Android malware. 

\subsection{Feature Extraction}

Three sensitive features of Android applications are considered in our approach, including permission, intent-action and API. Permission feature is our first selected feature because all the sensitive permissions that a malware need to operate on sensitive data must be declared explicitly in the AndroidManifest.xml file. Intent-action is mainly used to assist interaction and messaging between different components, which has been researched that it is highly related to sensitive operation\cite{Su2017A,Karbab2017Android}. API calls are the classes and methods that have been decompiled to smali grammar, which include low-level system calls. Hence, it's reasonable to take these three static features into consideration.
After extracting features, we vectorize the derived features and concatenate them to the final feature vector. 

\subsection{Selective Ensemble Method}

Selective ensemble method was first proposed by Zhou et al \cite{Zhou2002Ensembling}. 
An ensemble is a combination of predictions by multiple component learners like neural networks, which are trained for the same task. It is researched that an ensemble of neural networks can enhance the model's generalization ability\cite{Carney1999Confidence, Opitz1996Generating} and an ensemble of some of the component learners may be better than all\cite{Zhou2002Ensembling}. Genetic algorithm is introduced to pick out the best combination of component neural networks for ensemble. Here, we suppose that we have N component learners which is denoted as $f_i(i=1,2...N)$. Then $f_i (x)$ denotes the ith component learner's prediction on sample $x$. If the prediction is malicious, we denote it as $1$. If not, we denote it as $-1$. Bootstrap sampling method is adopted to generate the ensemble, which means we randomly sample the data with replacement for $N$ times. To denote the ensemble effect mathematically, we suppose a weight vector $\omega(\omega_1,\omega_2,\dots,\omega_N)$ that is $N$-dimensional with each dimension having the value of either $0$ or $1$. Then our selective ensemble combination predicted by component learners based on majority voting can be given as follows.
\begin{equation}
  H(x)=sgn(\sum_{i=1}^{N}\omega_if_i(x)) \label{hx}
\end{equation}

When $\omega_i$ equals $0$, the related component learner $f_i$ cannot make any effect on the final ensemble combination $H(x)$. And the component learners whose  $\omega_i$ is $1$ make the final ensemble prediction about whether the application is benign or malicious. To find the appropriate weight vector $\omega$, we introduce genetic algorithm to help us search the global optimal solution and obtain the best ensemble combination.
\subsection{Genetic Algorithm}

Genetic algorithm is adopted to obtain the best ensemble combination. The pseudocode of the genetic training algorithm of our approach is presented in Algorithm~\ref{alg1}.

\begin{algorithm}[htbp]
	\caption{Pseudocode of genetic algorithm}
	\label{alg1}
	\begin{algorithmic}[1]
		\STATE $population$ $\gets$ generate pop\_size chromosomes randomly
		\FOR{$iteration=1$ to $max\_iter$}
		\STATE crossover($population$)
		\STATE mutation($population$)
		\STATE evaluate the fitness function of population
		\STATE $new\_population$ $\gets$ select\_newpop($population$)
		\STATE $population$ $\gets$ $new\_population$
		\ENDFOR
	\end{algorithmic}
\end{algorithm}

\paragraph{Population Initialization}The weight vectors that are mentioned in \textit{Selective Ensemble Method} section are naturally the chromosomes owing to their inherent binary attributes. Hence, during this procedure we initialize the population by randomly generating multiple 0-1 vectors as chromosomes.

\paragraph{Fitness Function Evaluation}We consider two factors in our fitness function. One is ensemble accuracy, the other is the diversity of the component learners. It is researched that component learners with good accuracy and high diversity can achieve excellent performance than any of the component \cite{Choenni2000Design}.We define our fitness function F(x) as follows:
\begin{equation}
  F(x)=Accuracy\times D
\end{equation}
In equation (2), \emph{Accuracy} denotes the accuracy of an ensemble on the training data. \emph{D} is the diversity factor of an ensemble. Suppose \emph{i} and \emph{j} are separate component learners of ensemble \emph{E},  $x_k$ is the \emph{k}th sample of the data and $p_t(x_k)$ denotes the prediction that component learner \emph{t} makes on sample $x_k$. Then  we consider defining diversity factor \emph{D} using Euclidean distance:
\begin{equation}
  D(x) = \frac{1}{N}\sum_{i<j}\sqrt{\sum_{k=1}^{M}(p_i(x_k)-p_j(x_k))^2}
\end{equation}

In equation(3), \emph{N} denotes the number of the component learners of an ensemble. It can be seen that both the accuracy factor and diversity factor constrain the fitness function. If the predictions of component learners in an ensemble are prone to be the same, the diversity factor \emph{D} will decrease to 0, which make the fitness value so small to survive to the next generation. But if the predictions of the component learners are totally different, it will cause the accuracy factor decrease to 0, making the ensemble hard to survive. Only the ensembles with prediction difference on different part of the data and high accuracy are inclined to survive in the genetic algorithm.


\section{Preliminary Results \label{section4}}
In this section, we will elaborate the process of our experiment and prove the effectiveness of our approach in real world dataset.
\subsection{Datasets}

We collected a dataset that is composed of 16000 real world Android applications in total, 8000 for malwares and 8000 for benign ones. Malicious Apps are obtained from Information Security Lab of Peking University and each of them has been detected with multiple malicious actions using Android malware detection platform like \emph{VirusTotal}. Benign Apps are collected from Google Play, which have been manually checked. So we basically can ensure the correctness of our dataset. 
To simulate the circumstance in real world, we swap some proportion of malicious Apps and benign Apps, which is 10\%. To distinguish this dataset from the original one, we will refer this dataset as \emph{modified dataset}, and the other is referred as \emph{original dataset}. SEdroid will be evaluated on both datasets. In our experiment, we split our dataset into three parts, 60\% for training set, 20\% for validation set and 20\% for test set. For evaluation metrics, accuracy, precision rate, recall rate and F1-score are adopted to evaluate the model.

\subsection{Evaluation on Original Dataset}
Based on previous work, we consider SVM and DBN as component learners. As is shown in Table 1, when the dataset is correct or almostly correct, all the machine learning methods listed below can achieve good results on our original dataset. SVM is really useful, which achieves a 94.7\% precision rate and 94.1\% F1-score. DroidDeep\cite{Wang2017DroidDeepLearner} behaves even better, which has a 96.7\% precision rate and 95.2\% F1-score. For DBN and SVM, the selective ensemble method slightly enhance the performance at about 2\%, and the ensemble of DBN (SEdroid) performs even better than the ensemble of SVM, which is probably caused by the stronger representation ability of DBN. In general, all machine learning methods above performs good on the original dataset, and our approach has a slightly superiority than DroidDeep.
\begin{table}[htbp]
  \caption{Evaluation Results on Original Dataset\label{Table2}}
  \begin{center}
 
  \begin{tabular}{l|rrr} \hline
                  & Precision(M)  &Recall(M) &F1-score(M)  \\  \hline
     SVM           & 0.947         &0.936     &0.941        \\
     DroidDeep (DBN)  & 0.967         &0.937     &0.952  \\ \hline
     Ensemble of SVM &0.960         &0.972     &0.966   \\
     \textbf{SEdroid (DBN)}    &\textbf{0.979}         &\textbf{0.985}   &\textbf{0.982}       \\
       \hline
  \end{tabular}
  \end{center}
\end{table}

\subsection{Evaluation on Modified Dataset}

We did 30 repeated experiments to evaluate SEdroid compared to DroidDeep on the modified dataset
and the results are shown in Table~\ref{Table2}.
The performance of DroidDeep on modified dataset is not as good as on original dataset, which is caused by the data permutation in the modified dataset. The worst performance of DroidDeep is 89.7\% of precision rate and 90.3\% of recall rate. On the other hand, the best performance of DroidDeep can reach to 96.7\% of precision rate and 96.0\% of recall rate. This suggests that the performance of DroidDeep has a relatively large variance on the modified dataset, which can been seen in Fig.6 intuitively. In contrast, the performance of SEdroid has superior performance on modified dataset than DroidDeep, which has a 98.3\% precision rate and 98.1\% recall rate on average, which proves superiority and robustness of SEdroid.

\begin{table}[htbp]
  \caption{Evaluation Results on Modified Dataset\label{Table2}}
  \begin{center}
  
  \begin{tabular}{l|rrr} \hline
                  & Precision                  &Recall                         &F1-score  \\  \hline
     DroidDeep (worst)          & 0.897         &0.903     &0.900        \\
     DroidDeep (best)  & 0.967         &0.960     &0.963  \\ 
    DroidDeep (average)  & 0.938         &0.943     &0.940  \\  \hline 
     \textbf{SEdroid (average)}    &\textbf{0.983}         &\textbf{0.981}   &\textbf{0.982}       \\
       \hline
  \end{tabular}
  \end{center}
\end{table}

\section{Conclusion \label{section5}}
In this paper, we propose an Android malware detection approach called SEdroid, which is based on selective ensemble learning and genetic algorithm. 
SEdroid takes into account both accuracy and diversity of the ensemble of component learners, which increases the model's generalization ability and robustness.
In the evaluation, SEdroid managed to perform more effective and robust compared to other approaches.

\bibliography{references}

\begin{thebibliography}{1}

\bibitem{Sahs2012A}
Justin Sahs and Latifur Khan.
\newblock A machine learning approach to android malware detection.
\newblock In {\em Intelligence and Security Informatics Conference}, pages
  141--147, 2012.

\bibitem{Zhu2017DeepFlow}
Dali Zhu, Hao Jin, Ying Yang, Di~Wu, and Weiyi Chen.
\newblock Deepflow: Deep learning-based malware detection by mining android
  application for abnormal usage of sensitive data.
\newblock pages 438--443, 2017.

\bibitem{Su2017A}
Xin Su, Dafang Zhang, Wenjia Li, and Kai Zhao.
\newblock A deep learning approach to android malware feature learning and
  detection.
\newblock In {\em Trustcom/bigdatase/i​spa}, pages 244--251, 2017.

\bibitem{Karbab2017Android}
El~Mouatez~Billah Karbab, Mourad Debbabi, Abdelouahid Derhab, and Djedjiga
  Mouheb.
\newblock Android malware detection using deep learning on api method
  sequences.
\newblock 2017.

\bibitem{Zhou2002Ensembling}
Zhi~Hua Zhou, Jianxin Wu, and Wei Tang.
\newblock Ensembling neural networks: Many could be better than all.
\newblock In {\em ARTIFICIAL INTELLIGENCE}, 2002.

\bibitem{Carney1999Confidence}
J.~G Carney, P~Cunningham, and U~Bhagwan.
\newblock Confidence and prediction intervals for neural network ensembles.
\newblock In {\em International Joint Conference on Neural Networks}, pages
  1215--1218 vol.2, 1999.

\bibitem{Opitz1996Generating}
David~W. Opitz and Jude~W. Shavlik.
\newblock Generating accurate and diverse members of a neural-network ensemble.
\newblock {\em Advances in Neural Information Processing Systems}, 8:535--541,
  1996.

\bibitem{Choenni2000Design}
Sunil Choenni.
\newblock Design and implementation of a genetic-based algorithm for data
  mining.
\newblock In {\em VLDB 2000, Proceedings of International Conference on Very
  Large Data Bases, September 10-14, 2000, Cairo, Egypt}, pages 33--42, 2000.

\bibitem{Wang2017DroidDeepLearner}
Zi~Wang, Juecong Cai, Sihua Cheng, and Wenjia Li.
\newblock Droiddeeplearner: Identifying android malware using deep learning.
\newblock In {\em Sarnoff Symposium, 2016 IEEE}, pages 160--165, 2017.

\end{thebibliography}
\bibliographystyle{unsrt}

\end{document}